\documentclass[preprint,12pt]{elsarticle}

\usepackage{amssymb}

\usepackage[dvipsnames]{xcolor}
\usepackage[utf8]{inputenc}
\usepackage[T1]{fontenc}
\usepackage{array,multirow}
\usepackage{comment}

\usepackage{natbib}
\biboptions{sort&compress}

\usepackage{lineno}

\usepackage[
  letterpaper,
  left=3cm,
  right=3cm
]{geometry}

\journal{Indoor Air}

\begin{document}

\begin{frontmatter}



\title{Reconciling airborne disease transmission concerns with energy saving requirements: the potential of UV-C pathogen deactivation and air distribution optimization}



\cortext[cor]{Corresponding author}

\author[aff1]{Antoine Gaillard}
\ead{antoine0gaillard@gmail.com}
\author[aff2,aff3]{Detlef Lohse}
\author[aff1]{Daniel Bonn}
\author[aff4]{Fahmi  Yigit}
\ead{Fahmi.yigit@virobuster.nl}

\affiliation[aff1]{organization={Van der Waals-Zeeman Institute, University of Amsterdam}, 
            addressline={Science Park 904}, 
            city={Amsterdam},
            country={Netherlands}}

\affiliation[aff2]{organization={Physics of Fluids Group and Max Planck Center for Complex Fluid Dynamics, J. M. Burgers Centre for Fluid Dynamics, University of Twente}, 
            addressline={P.O. Box 217}, 
            city={Enschede},
            postcode={7500AE}, 
            country={Netherlands}}

\affiliation[aff3]{organization={Max Planck Institute for Dynamics and Self-Organization}, 
            addressline={Am Fassberg 17}, 
            city={37077 G\"ottingen},
            country={Germany}}

\affiliation[aff4]{organization={Virobuster International GmbH}, 
            addressline={K\"ohlershohner Stra{\ss}e 60}, 
            city={53578 Windhagen},
            country={Germany}}

\begin{abstract}
The COVID-19 pandemic caused a paradigm shift in our way of using heating, ventilation, and air-conditioning (HVAC) systems in buildings. In the early stages of the pandemic, it was indeed advised to reduce the reuse and thus the recirculation of indoor air to minimize the risk of contamination through inhalation of virus-laden aerosol particles emitted by humans when coughing, sneezing, speaking or breathing. However, such recommendations are not compatible with energy saving requirements stemming from climate change and energy price increase concerns, especially in winter and summer when the fraction of outdoor air supplied to the building needs to be significantly heated or cooled down. In this experimental study, we aim at providing low-cost and low-energy solutions to modify the ventilation strategies currently used in many buildings to reduce the risk of respiratory disease transmission. We find that ultraviolet germicidal irradiation (UVGI) modules added to the HVAC system are very efficient at inactivating pathogens present in aerosols, leading to good indoor air quality even with significant indoor air recirculation. Moreover, we show that an optimal placement of the air exhaust vents relative to air supply vents can improve the ventilation efficiency, with significant consequences in terms of energy savings.
\end{abstract}



\begin{keyword}
HVAC \sep COVID-19 \sep Virus \sep Aerosols \sep UVGI  




\end{keyword}

\end{frontmatter}


\section{Introduction}
\label{Introduction}

Heating, ventilation, and air-conditioning (HVAC) systems aim at providing safety against high CO$_2$ and volatile organic compound (VOC) concentrations, as well as comfortable temperature and relative humidity levels for, e.g., employees working in office buildings as well as students
and teachers in classrooms. With the COVID-19 pandemic, new hygienic concerns emerged and the importance of minimizing the risk of respiratory disease transmission in buildings was raised. Virus laden airborne aerosols, which are typically 1~$\mu$m in size and remain in suspension in the air after being produced by employees when speaking, coughing and sneezing 
\cite{bourouiba2020,miller2020,buonanno2020,somsen2020, abkarian2020,yang2020stone,rijn2020,bazant2021,bourouiba2021-arfm,chong2021,ng2021,bagheri2021,pohlker2021,poydenot2022,mutsch2022} have indeed been found to be one of the major transmission routes of the 
SARS-CoV-2 virus and other viruses
\cite{fisk2000health,li2007role,who2020ventilation,morawska2020a,morawska2020,jayaweera2020,prather2020,stadnytskyi2020,zhang2020,greenhalgh2021,cdc2021ventilation}. 
The risk of transmission therefore correlates with the rate at which indoor air is replaced by ‘clean’ (outdoor or decontaminated) air by the building’s ventilation system. While only outdoor air supply was advised during the COVID-19 pandemic, with minimum indoor air recirculation to avoid aerosol spreading and accumulation \cite{morawska2020a}, such strategy clashes with sustainability requirements since the heating or cooling of outdoor air represents a large fraction of the total energy consumption of buildings, which is over one third of society's global energy consumption \cite{morawska2021paradigm}. Therefore, other solutions must be explored to reduce the risk of respiratory disease transmission in an ecologically acceptable way.

Historically, the first HVAC systems had minimal recirculation and maximum supply of outdoor air, offering a good protection against potential virus-laden aerosols, as well as CO$_2$ and potential organic volatile compounds (VOCs). After the first oil crisis, more indoor air recirculation was needed to reduce energy costs. This led to health problems referred to as the `sick building syndrome' such as, for example, eye and upper respiratory symptoms attributed to high CO$_2$ concentrations \cite{tsai2012office}. Generic air filters such as F5/F7 are often introduced to improve indoor air quality, but to trap all aerosols finer HEPA filters are needed \cite{dreier2008transmission}. This suggests that, with insufficient air treatment, indoor air recirculation can lead to virus accumulation in buildings. Thermal wheels were therefore introduced in the 90's to minimize recirculation and increase the supply of fresh air from outside, which is now heated by the warm exhaust air from the inside without physical contact between the two air streams. However, in addition to their high construction costs and maintenance levels, these energy recovery heat exchangers are known to leak and send some indoor air back into the building (`carry-over' effect) \cite{simonson1998heat,enventus2018rotary,herath2020applicability}. Hence, assuming that even a small fraction of virus recirculation is problematic, thermal wheels may not be the solution to optimize protection against viruses.

A promising solution for optimizing the hygienic quality of indoor air while keeping the economical and ecological benefits of air recirculation lies in the deactivation of pathogens via chemical \cite{schwartz2020decontamination}, plasma \cite{filipic2020cold,lai2016evaluation} or short ultraviolet (UV-C) \cite{garcia2020back,hitchman2020new} air treatment. Ultraviolet germicidal irradiation (UVGI) has already proven effective to purify (drinking) water during the fight against tuberculosis, for which Niels Ryberg Finsen received the Nobel Prize in Medicine in 1903, as well as air in operation rooms \cite{hart1960bactericidal}. Since then, UVGI has been widely documented \cite{kowalski2009uvgi,menzies2003effect,sharp1939lethal,tseng2005inactivation} and was proven effective also against corona viruses \cite{walker2007effect,kowalski2020covid}. UVGI technologies are affordable, require low maintenance, and can be easily implemented as modules added to HVAC systems to treat recirculating indoor air before it is re-injected into the room. However, in spite of its great potential, experimental studies on the efficiency of this technology in the context of indoor air quality in buildings are still lacking to date. 

In addition to hygienic air treatment, the ventilation efficiency is key to reduce the risk of both short-range (via turbulent respiratory plumes emitted when, e.g., coughing or sneezing \cite{abkarian2020,chen2020short,bourouiba2014, smith20}) and long-range (after dilution and mixing in the room) disease transmission. In principle, long-range transmission can be reduced by increasing the ventilation power, which involves energy costs, or by improving the air distribution via wise placements of the air supply and exhaust vents to minimize pathogen concentrations in the breathing zone of occupants \cite{morawska2021paradigm,liu2022evaluation,el2022numerical,zhang2017numerical,cheong2021effect,abdolzadeh2019numerical,verma2017study}. In addition, future ventilation designs could rely on personalized ventilation concepts with clean air supplied near occupants \cite{morawska2021paradigm,xu2020effects}. However, a key question that, to our knowledge, has not been addressed to date is whether the ventilation efficiency, indexed by the CO$_2$ or aerosol elimination rate, for example, can be improved by modifying the air distribution strategy.

This study aims at providing further insight on finding a low-cost and ecological solution to modify the ventilation systems currently used in many buildings, in order to reduce the risk of respiratory disease transmission. To do so, we investigate the efficiency of UVGI air treatment in an office building with partly recirculated air, as well as the role of the air distribution strategy. We start by reiterating a simple model \cite{shair1974theoretical} correlating indoor CO$_2$ and aerosol concentrations with the ventilation efficiency in \S\ref{sec:model}. Our employed experimental methods are then presented in \S\ref{sec:materials_and_methods} and the results are presented in \S\ref{sec:results} and discussed in \S\ref{sec:conclusions}.

\section{Model}
\label{sec:model}

The room concentration of a gas or substance in suspension in the air, such as CO$_2$ or aerosols produced by humans (neglecting other sources) evolves over time and depends on the number of people present in the room and on the rate at which air is replaced by new (fresh or recirculated) air via the ventilation system. We recall here the simplest model for a well-mixed room \cite{shair1974theoretical} equivalent to a continuous stirred tank reactor \cite{davis2012fundamentals} which assumes a homogeneous room concentration and which is the basis of the well-known Wells–Riley model for airborne disease transmission \cite{wells1955airborne,riley1978airborne}. Because of filtration (via face masks \cite{shah2021experimental,howard2020,leung2020} or HEPA filters) and sedimentation-induced deposition on surfaces \cite{buonanno2020,bazant2021,yang2011dynamics,buonanno2020quantitative,smith20} affecting specifically airborne aerosols, the evolution in aerosol concentration may differ from that of CO$_2$ \cite{bazant2021monitoring,peng2021exhaled,stabile2021ventilation,pohle1992,rudnick2003risk}. These effects are neglected here for simplicity and aerosols and CO$_2$ are treated equally. We also neglect the natural \cite{yang2011dynamics,lin2019humidity,marr2019mechanistic,harper1961airborne} or active (via UVGI in our case) deactivation of pathogens contained in aerosols since we aim at describing the concentration of aerosols only, and not that of active pathogens themselves.

\begin{figure}
  \centerline{\includegraphics[scale=0.3]{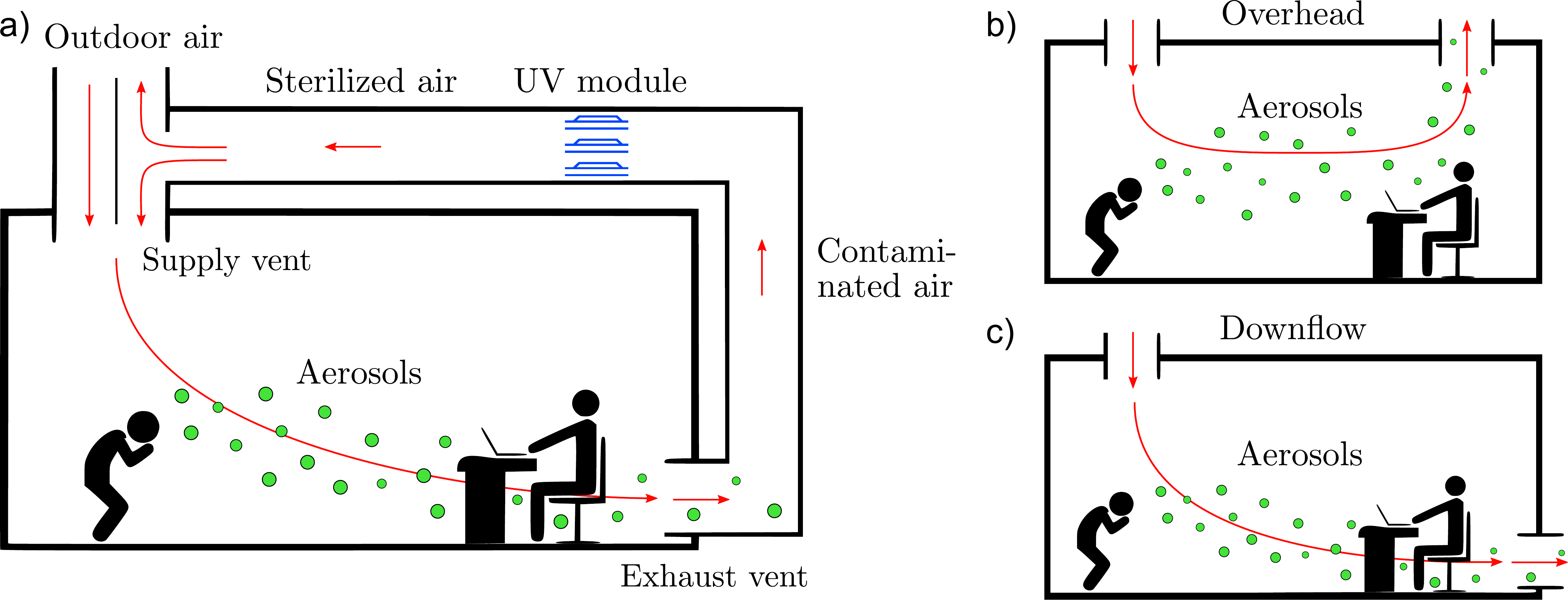}}
  \caption{(a) Sketch of the ventilation system. (b-c) Sketch or the air flow (arrows) in an overhead (b) and downflow (c) air distribution strategies. Two employees are sketched, one producing aerosols (circles) via coughing or sneezing, the other potentially breathing them. Note the specific case sketched here is not representative of all possible scenarios with, e.g., different positions of employees relative to supply and exhaust vents.}
\label{fig:sketch}
\end{figure}

The situation is described in figure \ref{fig:sketch}(a) where the air entering the room from supply vents consists of a fraction $x$ of fresh outdoor air and a fraction $1-x$ of recirculating indoor air captured by the exhaust vents before reentering the room after circulating through the AHU (Air Handling Unit). The concentration $C$ of either CO$_2$ or aerosols in a room of volume $V$ with $N$ people, each of them breathing at flow rate $q$, is set by the ventilation flow rate $Q$ via the differential equation 
\begin{equation}
    \frac{\mathrm{d}C}{\mathrm{d} t} = \lambda \left[ C_e - C + \alpha C_b \right]
    \label{eq:diff_eq_1}
\end{equation}
\noindent where $C_e$ and $C_b$ are the CO$_2$ or aerosol concentrations in the outdoor air and in human breath, respectively. Here we have introduced the air exchange rate
\begin{equation}
   \lambda = \frac{1}{\tau} = \frac{x Q}{V}
   \label{eq:lambda}
\end{equation}
\noindent and the ratio 
\begin{equation}
\alpha = N q/ x Q 
\end{equation}
of people vs. ventilation flow rate, which is typically $\alpha \ll 1$ in office buildings, as must be assumed to derive equation \ref{eq:diff_eq_1}. The general solution of equation \ref{eq:diff_eq_1} is 
\begin{equation}
    C(t) = C_{\infty} + (C_0 - C_{\infty}) e^{-\lambda t}
\end{equation}
\noindent where $C_0$ is the initial concentration at $t=0$ and where the concentration $C_{\infty}$ reached at times $t \gg \tau$ is given by
\begin{equation}
   C_{\infty} - C_e = \frac{N q C_b}{x Q} = \alpha C_b.
   \label{eq:Cinf}
\end{equation}
\noindent Equations \ref{eq:lambda} and \ref{eq:Cinf} show that both the equilibrium concentration difference $C_{\infty} - C_e$ and the characteristic time $\tau$ needed to reach it are inversely proportional to $xQ$ which, from a purely hygienic perspective, needs to be as high as possible to minimize the risk of airborne disease transmission. However, from an energy saving perspective, $x$ and $Q$ must remain reasonably low to minimize the energy cost of heating (in winter) or cooling down (in summer) the fresh outdoor air supplied to the room. 

\section{Materials and methods}
\label{sec:materials_and_methods}

Experiments are performed in a so-called open plan office building in Wierden, Netherlands, consisting of a 435~m$^2$ open space with ceiling height 2.6~m, where 20 to 40 employees from the company `Brand Builders', male and female aged between 18 and 70, work with continuous communication with each other, which may include speaking with raised voices. Employees do not have any obligation to wear a face mask and are asked to keep windows and doors shut, regardless of weather conditions.

While breathing or speaking, humans produce aerosols as well as CO$_2$ which, when other potential sources are negligible, are good indexers of the indoor air quality. Hence, the office room in Wierden is equipped with 12 CO$_2$ (MHZ19B) sensors and 5 air quality (SDS011) sensors placed on office desks and on the walls, either near the floor, at desk level or near the ceiling. The air quality sensors measure the concentration of particulate matter of size less than 2.5~$\mu$m (PM2.5) and less than 10~$\mu$m (PM10). Additionally, one CO$_2$ and one particle sensor are placed outside of the room to compare CO$_2$ and particle levels in the room with outdoor levels. We choose a time resolution of one data point per minute for all sensors.

Air is supplied to the room via 25 supply vents placed in the ceiling and exhausted via 15 exhaust vents placed either in the ceiling, via open traps of the dropped ceiling, or near the floor via tubes carrying the air above the dropped ceiling. These two different air distribution strategies, labeled `overhead' and `downflow',  respectively, are sketched in figure \ref{fig:sketch}(b,c). In the overhead strategy, where the tubes are closed, the ceiling traps placed above these tubes are opened while, in the downflow strategy, all ceiling traps are closed and the tubes are open. 

A direct measurement of the flow rate delivered by each supply vent was performed using an anemometer, yielding a value of the total ventilation flow rate of $Q_{max} = 6260$~m$^3$/h by summing over all supply vents. This value was measured at full ventilation capacity of the HVAC system. However, due to potential `shortcuts' between supply and exhaust vents, the effective air exchange rate $\lambda$ might be less than the optimal value $\lambda_{max} = xQ_{max}/V$ expected from an ideal ventilation efficiency (see model in \S\ref{sec:model}). This can be easily understood by considering that the CO$_2$ or aerosol concentration in the air exhausted via exhaust vents can indeed be lower than the average concentration across the room. Since the importance of these shortcuts depends on the position of the exhaust vents relative to the supply vents, different air distribution strategies are expected to yield different effective air exchange rates. In the following, results will be presented in terms of an effective flow rate $Q = \lambda V / x$ based on the measured CO$_2$ or aerosol elimination rate $\lambda$.

Two types of measurements are performed to estimate this effective available flow rate $Q$ for both overhead and downflow air distribution strategies. The first consists of continuous measurements the CO$_2$ and aerosol concentration in the room over several days between May and September 2022 and using the equilibrium concentration $C_{\infty}$ to estimate $Q$ for different air distribution strategies using equation \ref{eq:Cinf}, knowing the number of employees $N$ in the rooms every day. These measurements are performed at full ventilation capacity with $x=$30\% of supplied outdoor 
air.

The second and more straightforward approach consists in partially filling the room with either CO$_2$ or aerosols at a sufficiently high initial concentration $C_0$ before turning the ventilation on at full capacity with a fraction $x=100$\% of outdoor supplied air and without any employee in the room ($N=0$), in which case the concentration is expected to decrease exponentially from $C_0$ to $C_e$ at a rate $\lambda = Q/V$ from which $Q$ can be calculated, see the model in \S\ref{sec:model}. These experiments will be referred to as HADR (Hygienic Air Delivery Rate) measurements as they aim at measuring the effective rate $Q$ at which the air in a room is replaced by `fresh' air through the ventilation system. For CO$_2$ measurements, the room was filled with CO$_2$ by sublimating a few kilograms of dry ice, pouring warm water on it to accelerate the sublimation process. For aerosol measurements, the room was filled with aerosols by spraying a liquid with a special spray nozzle provided by Medspray\textsuperscript{\textregistered}. The size distribution of droplets (before evaporation) and aerosols (after evaporation) was measured by a laser diffraction technique provided by Malvern Panalytical\textsuperscript{\textregistered} (Spraytec). The average droplet and aerosol diameters are 7 and 4~$\mu$m, respectively, when spraying at a flow rate of 34~ml/h, from which we estimate that $3 \cdot 10^9$ droplets/aerosols are produced per minute. The sprayed liquid consists of water with 1~wt\% glycerin and 0.5wt\% NaCl so that, after water has evaporated, the glycerin and NaCl core remains in the form of an aerosol particle. Fans were used to homogenize the concentration during dry ice sublimation and aerosol spraying and were turned off before starting the experiments, i.e., before turning the ventilation system on. CO$_2$ and aerosol measurements were performed on November 12th, 2022, and September 3rd, 2022, respectively, both overhead and downflow air distribution strategies being tested on the same day.

The building's HVAC system, which doesn't include a thermal wheel or HEPA filters, is equipped with an UltraViolet Germicidal Irradiation (UVGI) system provided by Virobuster\textsuperscript{\textregistered} placed before the outdoor air injection point so that recirculating indoor air is treated before re-entering the room, as sketched in figure \ref{fig:sketch}(a). The UVGI module is about 1~m long and exposes the circulating air to UV-C light of wavelength 254~nm with an intensity around 600~J/m$^2$. To test the efficiency of this technology on the air quality, air samples were collected every week between May and September 2022 by an certified validation company at five different locations across the room, the same locations every week, at heights 0.4~m, 0.75~m, and 2~m from the floor. The air sampling protocol consists of creating a 6~m$^3$/h air flow through a Sartorius gelatin filter with 0.3~$\mu$m pores for 10 minutes during which employees were asked to not approach the sampling locations. Two additional air samples were collected, one outside to compare indoor air quality with outdoor, and one in the air handling unit, after the UVGI module and before the outdoor air injection point, to check whether the air leaving the module is significantly affected by the UV-C treatment. The results for every sample are expressed in terms of a number of colony forming units (CFU), including both bacteria and fungi, the two being easily distinguishable. Measurements are divided in time periods of at least two weeks corresponding to either overhead or downflow air distribution strategies, each strategy being tested with UVGI either turned on or turned off at full ventilation capacity with $x=30$\% of outdoor supplied air.

\section{Results}
\label{sec:results}

\subsection{HADR experiments}

\begin{figure}
  \centerline{\includegraphics[scale=0.54]{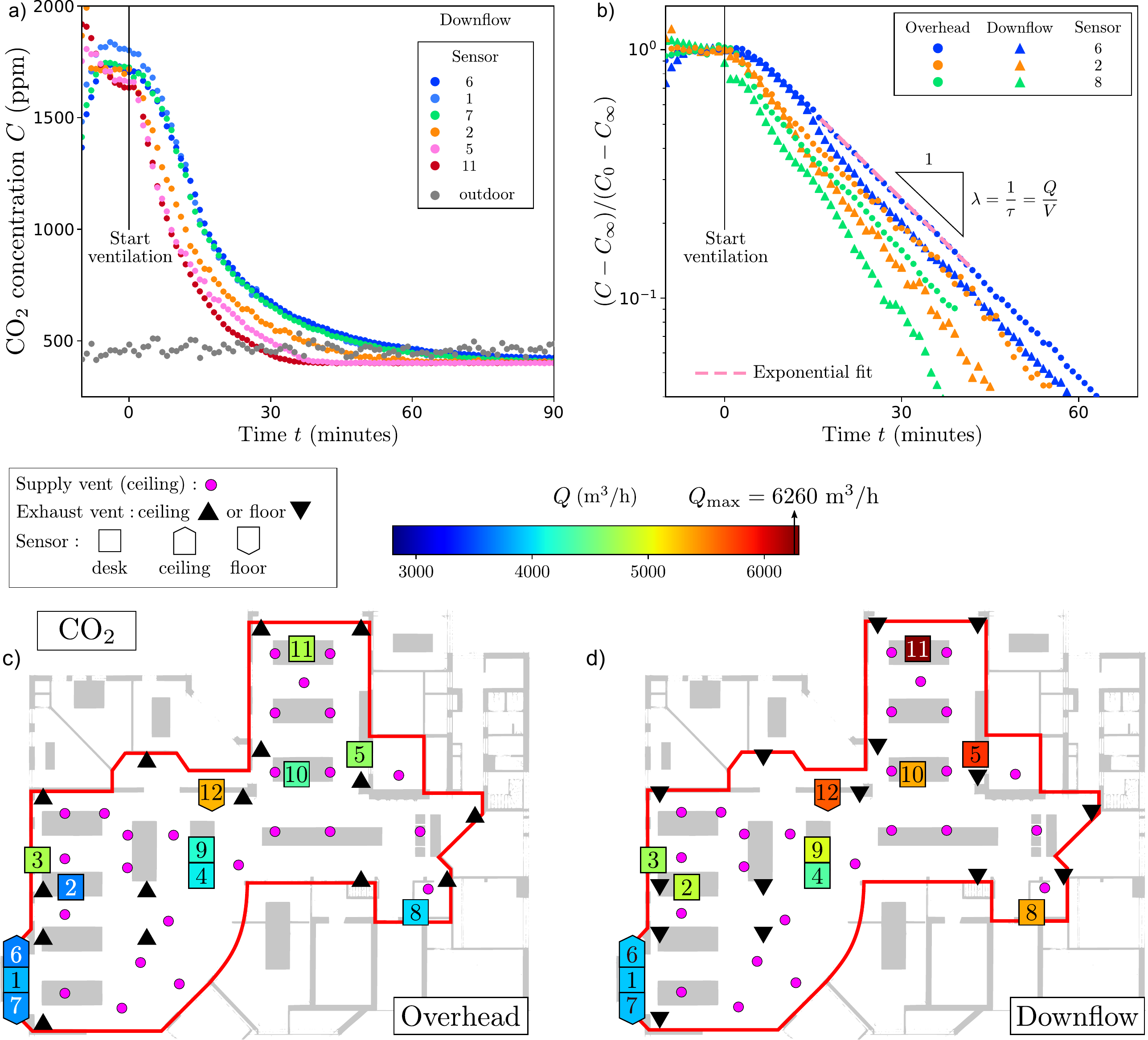}}
  \caption{(a) Time evolution of the CO$_2$ concentration in a HADR experiment where the room is initially filled with CO$_2$ by sublimating dry ice before turning the ventilation on at $t=0$ for different sensors in a downflow air distribution strategy. (b) Comparison of the re-normalized concentration decay curves $(C-C_{\infty})/(C_0 - C_{\infty})$ used to measure the air exchange rate $\lambda = Q/V$ for overhead and downflow air distribution strategies for three sensors. (c,d) Map of the effective flow rate $Q$ measured for all sensors in overhead (c) and downflow (d) air distribution strategies. The zone outlined in red is the room where experiments were carried out, all doors leading to other rooms were closed. The position of supply vents in the ceiling, exhaust vents in the ceiling (c) or near the floor (d) and sensors at desk level or near the ceiling or floor are shown by symbols presented in the legend. The sensor symbols are colored based on their $Q$ value following a code from blue (low $Q$) to red (high $Q$).}
\label{fig:HADR_CO2}
\end{figure}

The results of HADR measurements for CO$_2$ are presented in figure \ref{fig:HADR_CO2}(a) where, starting from a high initial CO$_2$ concentration $C_0$ of about $1700$~ppm, the concentration drops down to equilibrium values $C_{\infty}$ around $400$~ppm, close to the outdoor concentration. The rate at which the concentration decreases varies among different sensors placed at different locations across the room. Re-normalized decay curves $(C-C_{\infty})/(C_0 - C_{\infty})$ are compared in figure \ref{fig:HADR_CO2}(b) for three sensors for both overhead and downflow air distribution strategies. We find that (i) the decay is exponential, in agreement with the model presented in \S\ref{sec:model}, allowing measurement of the air exchange rate $\lambda = Q/V$ ($x=1$ here since there is no indoor air recirculation) from which the effective flow rate $Q$ is measured for each sensor, and that (ii) this effective flow rate is generally larger in downflow than in overhead. 

This is further demonstrated in the maps of figure \ref{fig:HADR_CO2}(c,d), showing the position of supply and exhaust vents in the room as well as the position of CO$_2$ sensors, which are colored based on their respective $Q$ value in overhead (c) and downflow (d) air distribution strategies. We observe that the $Q$ value measured for almost every sensor is larger in downflow than in overhead. The maps also reveal inhomogeneities, with for example a less well ventilated region at the bottom left, with no significant differences between sensors placed at different heights. 

Averaging over all sensors leads to effective flow rates $\langle Q \rangle = 4260$~m$^3$/h in overhead and $4930$~m$^3$/h in downflow with a standard variation of about $600$~m$^3$/h in both cases. These values are respectively 32\% and 21\% smaller than the maximum flow rate $Q_{max}$ that would be reached in the absence of shortcuts in the air flow, meaning that adopting a downflow air distribution strategy reduces the importance of these shortcuts by improving the effective average available flow rate $\langle Q \rangle$ by 16\% compared to overhead.

\begin{figure}
  \centerline{\includegraphics[scale=0.54]{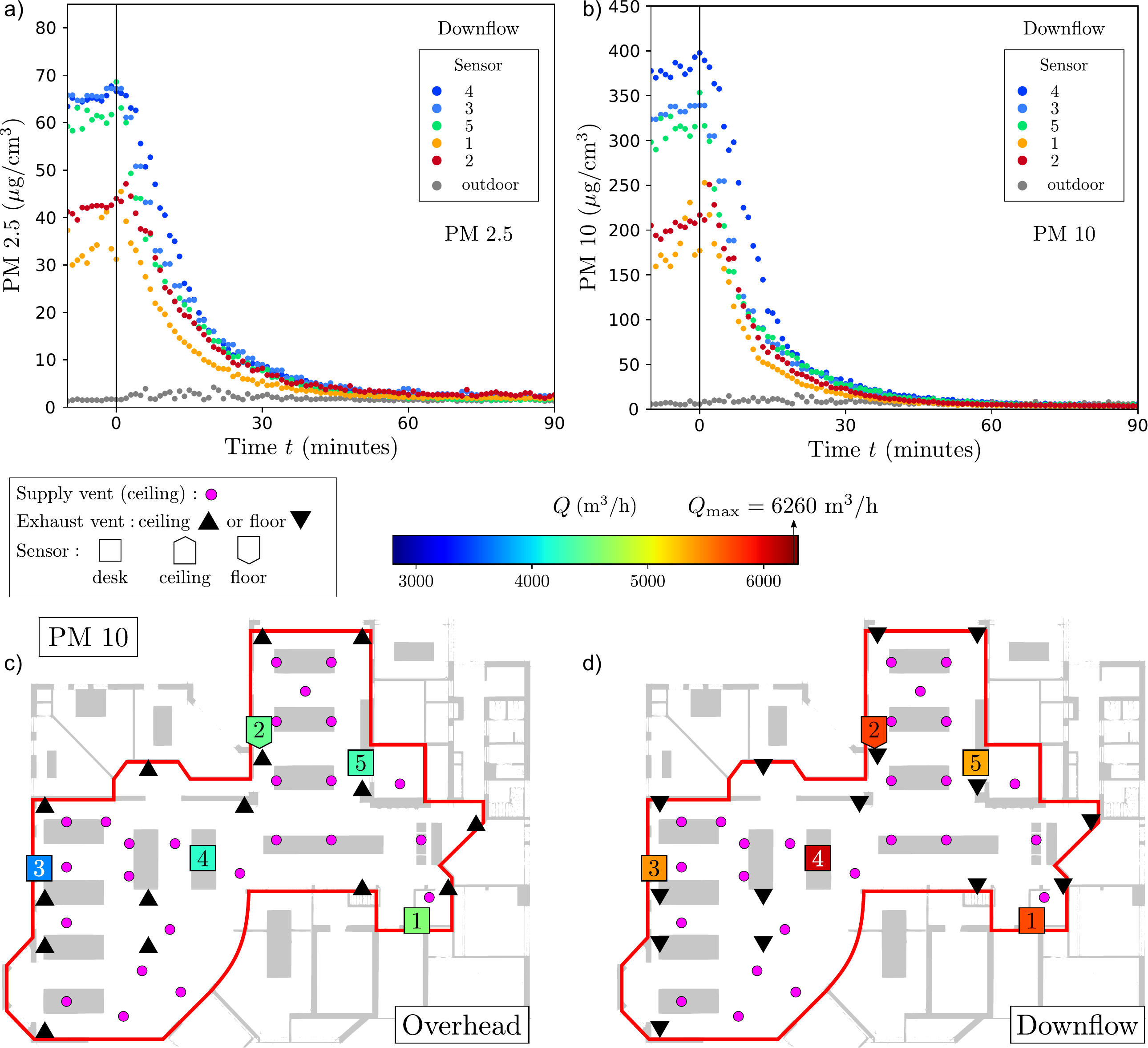}}
  \caption{(a,b) Time evolution of the concentration of aerosols of size less than 2.5~$\mu$m (PM2.5) (a) and 10~$\mu$m (PM10) (b) in a HADR experiment where the room is initially filled with aerosols of different sizes via a special spraying technique before turning the ventilation on at $t=0$ for different sensors in a downflow air distribution strategy. (c,d) Map of the effective flow rate $Q$ measured for all sensors in overhead (c) and downflow (d) air distribution strategies for aerosol of size less than 10~$\mu$m (PM10). The zone outlined in red is the room where experiments were carried out, all doors leading to other rooms were closed. The position of supply vents in the ceiling, exhaust vents in the ceiling (c) or near the floor (d) and sensors at desk level or near the ceiling or floor are shown by symbols presented in the legend. The sensor symbols are colored based on their $Q$ value following a code from blue (low $Q$) to red (high $Q$).}
\label{fig:HADR_aerosols}
\end{figure}

Similar results are found for aerosol HADR measurements, as shown in figures \ref{fig:HADR_aerosols}(a,b), showing the decay curves of the concentration in aerosols of size less than 2.5~$\mu$m (PM2.5) (a) and 10~$\mu$m (PM10) (b) in a downflow air distribution strategy. Note that the initial concentration $C_0$ reached before turning the ventilation on is larger for PM10 than for PM2.5, meaning that the spraying technique used to fill the room with aerosols produces more particles of size larger than 2.5~$\mu$m. As shown in the maps of figures \ref{fig:HADR_aerosols}(c,d), like for CO$_2$, the effective flow rate $Q$ calculated from these exponential decay curves is larger in downflow than in overhead for all sensors. Averaging over all sensors leads to effective flow rates $\langle Q \rangle = 4180$~m$^3$/h in overhead and $5390$~m$^3$/h in downflow for PM2.5 and $\langle Q \rangle = 4240$~m$^3$/h in overhead and $5660$~m$^3$/h in downflow for PM10, with standard variations of about $400$~m$^3$/h in all four cases. This means that adopting a downflow air distribution strategy improves the effective flow rate by about 30\% for aerosols, compared to overhead, which is a more pronounced improvement than for CO$_2$. 

Flow rates and improvement percentages are presented in table~\ref{tab:Q}. The more pronounced improvement observed for aerosols compared to CO$_2$ suggests that, in addition to the reduction in air shortcuts between supply and exhaust vents affecting both CO$_2$ and aerosols, a downflow air distribution strategy also enhances the elimination rate of aerosols via a second aerosol-specific mechanism. A possible interpretation is that, when being forced towards the floor, aerosols are more likely to be deposited on surfaces as they encounter more obstacles such as tables, chairs and other furniture, which would also occur in the absence of ventilation due to gravitational sedimentation \cite{smith20}.

\subsection{Continuous concentration measurements}

\begin{figure}
  \centerline{\includegraphics[scale=0.5]{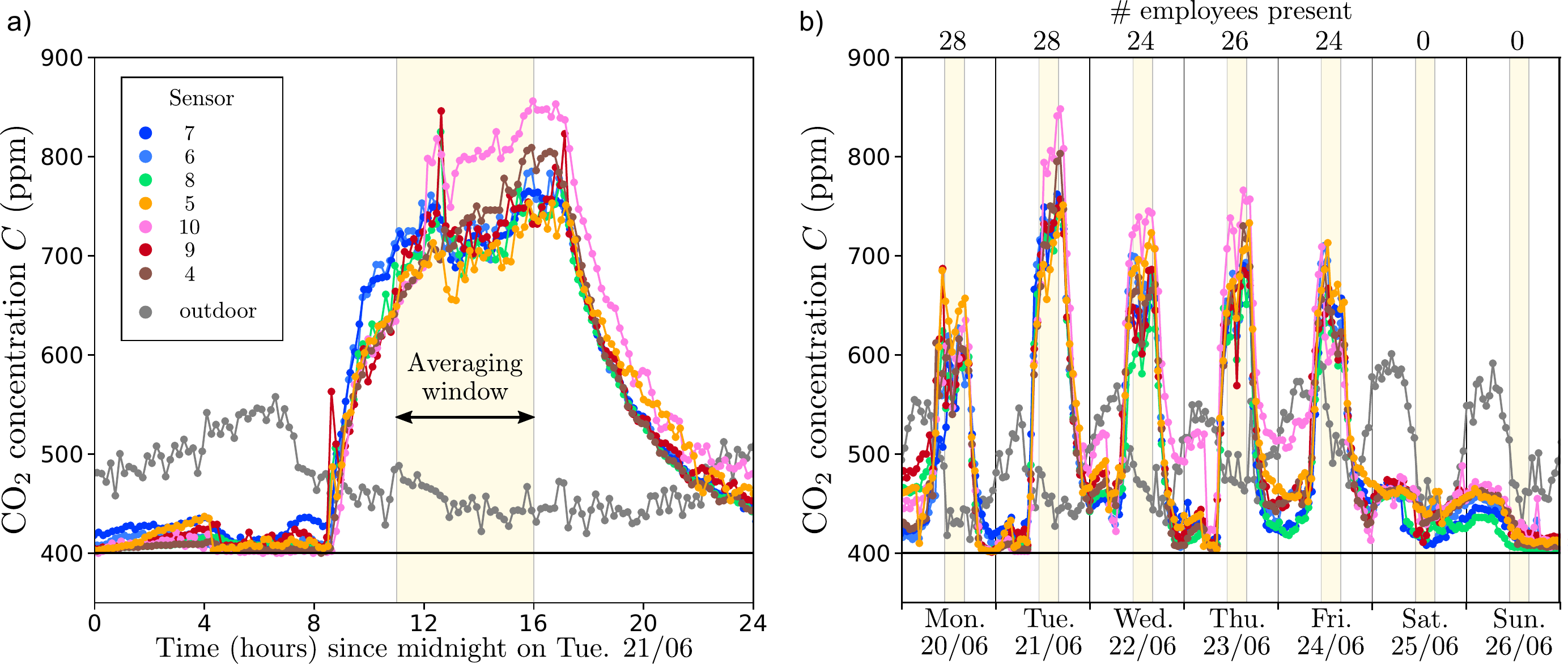}}
  \caption{Time evolution of the CO$_2$ concentration in the room for different sensors over one working day (a), Tuesday June 21st and one full week (b) from Monday June 20th to Sunday June 26th, with the number of employees present in the room indicated in the top $x$-axis. The equilibrium concentration $C_{\infty}$ is calculated by averaging over all indoor sensors between 11am and 4pm. While the real time resolution is one data points per minute, we only show enough points to see the trend in this figure. The air distribution strategy in place during these days was ``overhead''.}
\label{fig:continuous_CO2}
\end{figure}

The time evolution of the CO$_2$ concentration during normal working days with employees present in the room is shown in figure \ref{fig:continuous_CO2} over one day (a) and over one week (b) for an overhead air distribution strategy. The indoor concentration starts increasing at around 8am and start decreasing at around 5pm, consistent with typical working hours, while the outdoor CO$_2$ concentration increases slightly during the night and decreases in the morning, consistent with the plant breathing cycle (CO$_2$ release during the night and captures during the day). We can therefore correlate the indoor CO$_2$ concentration with the presence of people. However, as shown in figure \ref{fig:continuous_CO2}(b) where the number of employees present each day is indicated on the top $x$-axis, the indoor CO$_2$ concentration doesn't correlate perfectly with the number of employees present, probably due to some employees partially working outside of the main room where the concentration was measured (i.e., outside of the region outlined in red in figures \ref{fig:HADR_CO2}(c,d) and \ref{fig:HADR_aerosols}(c,d)). For that reason, estimating the effective flow rate for the two air distribution strategies using equation \ref{eq:Cinf}, averaging over several days between 11am and 4pm for each strategy, yields significant standard deviations. Using $C_b = 40,000$~ppm and $q = 220$~l/h gives average values $\langle Q \rangle = 4260$~m$^3$/h in overhead and $4670$~m$^3$/h in downflow with standard variations of about $1300$~m$^3$/h in both cases, which is larger than the 10\% improvement of downflow compared to overhead estimated from these values, shown in table \ref{tab:Q}. We therefore believe that HADR measurements are a more reliable way to measure the effective flow rate and to detect differences between different air distribution strategies.

We measured that, contrary to CO$_2$ concentrations, there was no difference in aerosol concentration between weekend and working days, meaning that too few aerosols were generated by human activity to be detected by our sensors. Hence, no effective flow rate could be estimated from continuous aerosol concentration measurements.

\begin{table}
\setlength{\tabcolsep}{6pt}
  \begin{center}
  \begin{footnotesize} 
\def~{\hphantom{0}}
  \begin{tabular}{ccccc}
    \hline
                                    & HADR CO$_2$  & Cont. CO$_2$ & HADR PM2.5 & HADR PM10 \\[8pt]
     $Q_{max}$ (m$^3$/h) - Optimal  & 6260         & 6260         & 6260       & 6260 \\[8pt]
     $Q$ (m$^3$/h) - (O)verhead     & 4260         & 4260         & 4180       & 4240 \\[8pt]
     $Q$ (m$^3$/h) - (D)ownflow     & 4930         & 4670         & 5390       & 5660 \\[8pt]
     \% Improvement O to D          & 16           & 10           & 29         & 33 
  \end{tabular}
  \caption{Effective flow rates $Q$ measured from HADR measurements (no employees in the room) with CO$_2$ and aerosols (PM2.5 and PM10) and from continuous CO$_2$ measurements (employees in the room) for overhead and downflow air distribution strategies. The last line shows the improvement of downflow compared to overhead in percentage. In the absence of shortcuts, the measured flow rate should be equal to the optimal value $Q_{max} = 6260$~m$^3$/h measured with an anemometer.} 
  \label{tab:Q}
  \end{footnotesize} 
  \end{center}
\end{table}

\subsection{Effect of ultraviolet germicidal irradiation}

\begin{figure}
  \centerline{\includegraphics[scale=1]{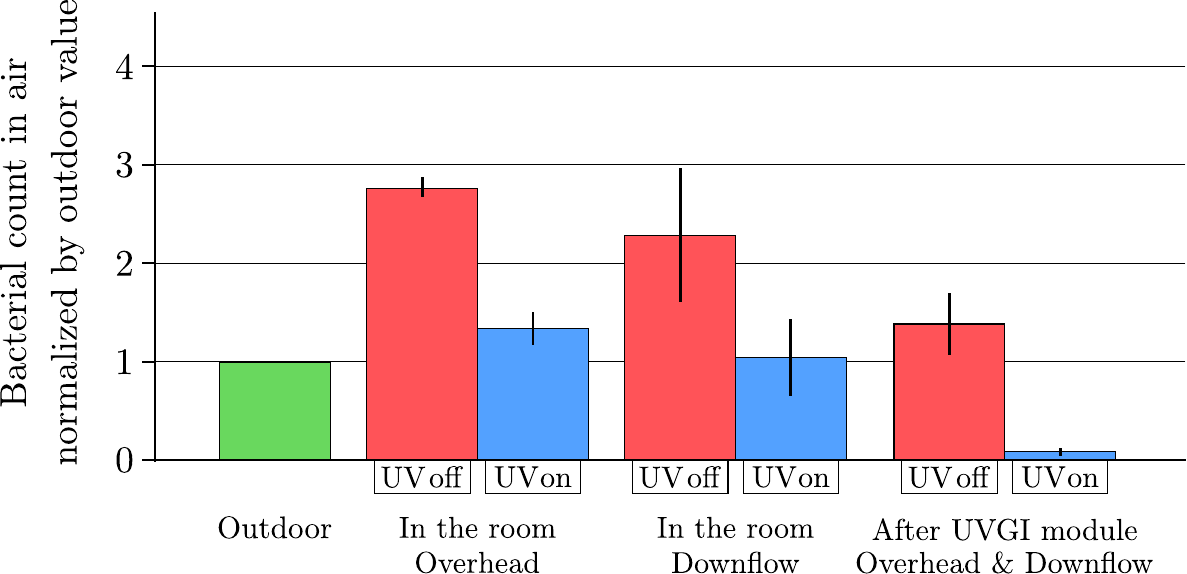}}
  \caption{Bacterial count in air sampling, normalized by the outdoor value, outdoor (green), indoor (averaged over different locations in the room) with employees present and in the air handling unit, after the UVGI module, with ultraviolet germicidal irradiation either turned off (red) or on (blue). The data is averaged over different weeks where, for values in the room, we distinguish between weeks with overhead and downflow air distribution strategies. Error bars correspond to the standard variation over different weeks.}
\label{fig:CFU}
\end{figure}

The air quality measurements carried out on Friday every week showed strong variations of both outdoor and indoor air quality between different weeks, with numbers of bacterial colony forming units (CFU) ranging typically between 10 and 1000 for outdoor air samples. This is not surprising since the presence of bacteria in the air is expected to depend on, e.g., the temperature and relative humidity \cite{hoeksma2015effect,qiu2022effects}. Days of high outdoor bacterial concentration also showed high indoor bacterial concentration when the UVGI module was off. Hence, for every measurement day, we normalize the indoor air quality (quantified by the number of bacterial CFU) measured at different locations in the room and in the air handling unit by the outdoor air quality. Strong air quality variations are observed for different air sampling locations in the room for a given day, the location of  the worst measured air quality not being the same every week. This is not surprising since the amount of bacteria in the air at a given location can vary significantly depending on the presence of nearby sources such as plants or food whose location can vary from day to day. Hence, we average over all five locations in the room for every measurement day to obtain an average room air quality.

The final results are presented in figure \ref{fig:CFU} where the air quality in the room and after the UVGI module is compared to the outdoor air quality. We average over weeks corresponding to the same air distribution strategy (overhead or downflow) for the air quality in the room and over all weeks for the air quality after the UVGI module, distinguishing in both cases between weeks where the UVGI was turned on (blue) and weeks where it was turned off (red). The error bars represent the variation over different weeks. These results show a strong improvement of the air quality in the room when UVGI is turned on, typically by a factor two, making indoor air almost as clean as outdoor. This is confirmed by the sharp reduction in bacterial count after the UVGI module when it is turned on. No significant difference is measured between the two air distribution strategies.

\section{Conclusions and discussions}
\label{sec:conclusions}

Our experiments confirm that ultraviolet germicidal irradiation (UVGI) technologies can significantly improve the indoor air quality in buildings when indoor air is partly re-injected in the room via recirculation, bringing the level of active airborne bacteria almost as low as in the outdoor air. Since viruses are much weaker than bacteria, and since the UV-C intensity of around 600 J/m$^2$ of the UVGI technology used in this study surpasses the required 90\% inactivation energy of SARS-CoV-2 (27~J/m$^2$ \cite{kowalski2020covid}), it can be assumed that the level of active airborne viruses will also be significantly reduced after UVGI treatment. Hence, assuming a virus-free outdoor air, the risk of airborne contamination is significantly reduced. We therefore conclude that energy saving requirements, from which air recirculation is encouraged, can be reconciled with hygienic requirements when properly treating the air re-injected in the building.


The effective flow rate $Q$, or equivalently the effective air exchange rate $\lambda$, available to replace indoor air with `clean' (outdoor or treated indoor) air was shown to be systematically less than the flow rate expected from direct measurements of supply vent delivery. This is caused by `shortcuts' in the global air flow of the room between supply and exhaust vents, which cause the air being exhausted to be more similar to the clean supplied air than to the average air in the room, either in terms of CO$_2$ or aerosol concentration or, consequently, in terms of temperature and humidity. While air quality measurements did not allow us to rank the two different air distribution strategies tested (overhead and downflow), a clear improvement of up to 30\% was measured in the effective air flow rate $Q$ for downflow compared to overhead, proving that supply-exhaust air shortcuts can be significantly minimized by changing the air distribution strategy. Future work is needed to characterize the optimal position of inlet and exhaust vents to minimize these shortcuts and to compare a downflow situation to the reverse case (not considered in the present study) where air is supplied from the floor and exhausted from the ceiling.


The improvement in ventilation capacity implies that, in addition to a better protection against virus-laden aerosols, a better air distribution strategy is also beneficial in terms of energy savings. Indeed, in winter for example, the presence of supply-exhaust shortcuts means that a fraction of the precious warm air supplied to the room is directly thrown away into exhaust vents before properly mixing in the room. Hence, when limiting these shortcuts, less ventilation energy is needed to keep the indoor air temperature at a prescribed value, which has a significant economic impact. In other words, an optimal air distribution strategy can lead to the same indoor air quality as for another (less optimal) strategy, both in terms of temperature and aerosol safety, with less ventilation power. For example, on a 4$^{\circ}$C winter day with 100~\% outdoor air supply (no indoor air recirculation), the thermal power needed to warm the outdoor air flowing in the air handling unit at flow rate $Q_{max}= 6260$~m$^3$/h up to 21$^{\circ}$C is 35~kW based on the density and specific heat capacity of air. Once in the room, our effective flow rate measurements in ``overhead'' showed that about 30\% of the flow rate is lost due to shortcuts, implying a power loss of about 11~kW. In contrast, in ``downflow'', based on CO$_2$ and aerosol measurements, only 20\% to 10\% of the flow rate is lost, respectively, implying a power loss of only 7.4~kW to 3.5~kW.

We note that the use of UVGI modules in HVAC systems could, in principle, allow the use of rotary heat wheels, since the deactivation of pathogens that, without treatment, would be re-injected in the building via the `carry-over effect', could ensure an acceptable indoor air quality. Moreover, in contrast to air filters, UVGI modules can be easily switched on when required and don't introduce any additional pressure drop in the air handling unit (AHU). This solution is therefore not only ecologically justifiable, but also economically viable. In principle, UVGI modules also allow the use of cheap and small AHU without separated exhaust and supply air, i.e., without rotary heat wheels. This is particularly interesting since these ventilation systems allow, via indoor air recirculation, an easier regulation of the indoor air humidity, especially in winter, with the associated benefice for occupants' well-being.\\

\textbf{Conflicts of interest} 

The authors declare no conflicts of interest.\\

\textbf{Funding} 

The authors acknowledge funding by NWO (Dutch Research Council) under the MIST (MItigation STrategies for Airborne Infection Control) program.\\

\textbf{Acknowledgments}

The authors would like to thank Sander G. Huisman for fruitful discussions.


\begin{small}
\bibliographystyle{elsarticle-num} 
\bibliography{2ph_literatur.bib}
\end{small}




\end{document}